\documentclass[prl,letterpaper,aps,10pt,superscriptaddress,twocolumn,floatfix,showpacs]{revtex4-1}
\usepackage{amsmath,graphicx,amssymb,braket,xcolor,subfigure,upgreek}

\begin{document}

\title{Protected state enhanced quantum metrology with interacting two-level ensembles}

\author{Laurin Ostermann}
\author{Helmut Ritsch}
\affiliation{Institut f\"ur Theoretische Physik, Universit\"at
Innsbruck, Technikerstrasse 25, A-6020 Innsbruck, Austria}

\author{Claudiu Genes}
\affiliation{Institut f\"ur Theoretische Physik, Universit\"at
Innsbruck, Technikerstrasse 25, A-6020 Innsbruck, Austria}

\affiliation{ISIS (UMR 7006) and IPCMS (UMR 7504), Universit\'{e} de
Strasbourg and CNRS, Strasbourg, France}

\date{\today}

\begin{abstract}
Ramsey interferometry is routinely used in quantum metrology for the
most sensitive measurements of optical clock frequencies.
Spontaneous decay to the electromagnetic vacuum ultimately limits
the interrogation time and thus sets a lower bound to the optimal
frequency sensitivity. In dense ensembles of two-level systems the
presence of collective effects such as superradiance and
dipole-dipole interaction tends to decrease the sensitivity even
further. We show that by a redesign of the Ramsey-pulse sequence to
include different rotations of individual spins that effectively
fold the collective state onto a state close to the center of the
Bloch sphere, partial protection from collective decoherence and
dephasing is possible. This allows a significant improvement in the
sensitivity limit of a clock transition detection scheme over the
conventional Ramsey method for interacting systems and even for
non-interacting decaying atoms.
\end{abstract}


\maketitle

\textbf{\emph {Introduction}} The precise measurement of time using
suitable atomic transitions is a major achievement of quantum
metrology. The Ramsey interferometry procedure plays a crucial role
in this context as it allows a quite accurate locking of the
microwave or optical oscillator to the transition frequency in the
atom. Typical early realizations were based on an atomic beam or a
laser-cooled atomic fountain later on~\cite{wynands2005atomic},
where the atoms would interact with two consecutive Rabi pulses.
With optical lattices (details see~\cite{bloch2005ultracold}) and
proposals for optical lattice clocks, e.g.
~\cite{takamoto2005optical} time measurements were expected to
become even more accurate due to longer interaction times and  the
elimination of collisions (see
~\cite{oates2012optical,lemonde2009optical} for recent reviews). To
reduce quantum projection noise (scaling as $1/\sqrt{N}$, where $N$
is the number of atoms) and to speed up the measurement,
experimental setups usually involve an as  large as possible number
of atoms. In a finite volume, of course, this brings collective
effects like superradiance and dipole-dipole shifts to the table
~\cite{ostermann2012cascaded}. While some techniques rely on the
engineering of particular geometries without the need to alter the
internal atomic states \cite{chang2004controlling}, exploiting the
uncertainty principle by employing squeezed states
~\cite{oblak2005quantum,wineland1992spin,meiser2008spin} can be
helpful as well to achieve less noise with lower atom
numbers~\cite{leroux2010focus, leroux2010orientation}. These
techniques heavily rely on entanglement
\cite{louchet2010entanglement, borregaard2013near} among atoms
and require very careful preparation and isolation of the ensemble.

When, finally, interrogation times reached the lifetime of the
excited state, spontaneous emission became a critical factor for the
contrast of the Ramsey fringes. Interestingly, despite the use of
long lived clock states, for multiple atoms in close proximity to
each other, collective spontaneous emission can still reach a
detrimental magnitude. Here, the common atomic coupling to the same
electromagnetic vacuum fluctuations enhances spontaneous emission by
a factor proportional to the atom
number~\cite{lehmberg1970radiation,ficek2002entangled}. While this
is usually limited to volumes of the order of a cubic wavelength, in
regular arrays, such as an optical lattice, the effect can extend
over tens of lattice sites~\cite{zoubi2013excitons}. In addition,
excitonic energy level shifts in lattices can also induce
significant dephasing of the Ramsey signal, which cannot be removed
by simple echo techniques.

In this paper, we propose a strategy that works on the level of the
Ramsey pulses and which we dub the 'asymmetric Ramsey technique', in
contrast to the typical symmetric Ramsey technique that employs only
identical $\pi/2$ pulses applied to all atoms. While the
conventional Ramsey technique excites superposition states, which
possess a maximum dipole moment and thus are most sensitive to
superradiance, this new approach allows the selection of long-lived
collective states (or 'dark states') to improve the sensitivity of
the clock signal. The procedure requires two further manipulations
of the atomic dipoles in addition to the usual sequence: after the
initial $\pi/2$ pulse is applied, each atomic coherence is rotated
by a distinct phase, resulting in a subradiant collective state
(with a lifetime which can be even longer than that of the
independent atoms). This results in a state of vanishing classical
collective dipole which is typically well-protected from the
environmentally-induced decoherence~\cite{zoubi2010metastability}.
As a clarifying example, in the Dicke
limit~\cite{dicke1954coherence} of atoms positioned at the same
spot, a state that exhibits infinite lifetime exists and is
therefore perfectly suitable for state-protective spectroscopy.

\begin{figure*}[t]
 \includegraphics[width=1.99\columnwidth]{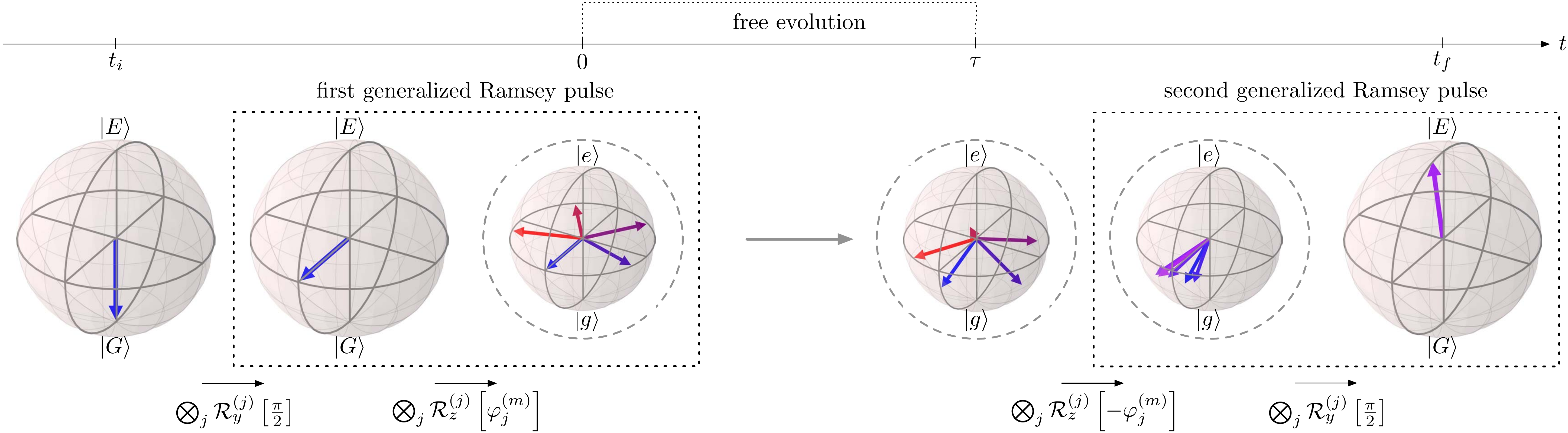}
 \caption{\emph{State protective Ramsey sequence.} The ensemble of $N$ spins
is prepared with all spins down in a collective coherent pure spin
state on the surface of the collective Bloch sphere (radius $N/2$).
Individual $\pi/2$ pulses are followed by phase encoding operations
of angles $\varphi_{j}^{(m)}=(2\pi m/N)(j-1)$ where $j=1,...N$ and
$m=1,...[N/2]$, which corresponds to bringing the total spin to a
zone close to the center of the Bloch sphere (notice that the third,
fourth and fifth steps are shown on the small Bloch spheres of
radius $1/2$ corresponding to single system states). After
interrogation time $\tau$, the phase encoding operation is reversed
and the second set of individual $\pi/2$ pulses prepare the ensemble
(now in a mixed state shown on the large collective Bloch sphere)
for the detection of the population difference signal.}
 \label{fig1}
\end{figure*}

\textbf{\emph {Model for N atoms}} We assume a collection of $N$
identical two-level emitters with levels $\left \vert
g\right\rangle$ and $\left\vert e \right\rangle$ separated by
$\omega_0$ in a general geometry defined by the positions
$\mathbf{r}_i$ for $i=1,...N$ and the angle $\theta$ drawn between
the (identical) transition dipoles and their separations. We define
individual Pauli ladder operators $ \sigma _i^\pm$ and subsequently
$\sigma_i^x=\sigma_i^+ + \sigma_i^-$, $\sigma_i^y=-i(\sigma_i^+
-\sigma_i^-) $ and $\sigma_i^z= \sigma_i^+ \sigma_i^- - \sigma_i^-
\sigma_i^+$ as well as the corresponding unitary rotations
$\mathcal{R}^{(j)}_\mu [\varphi] = \exp \left( i \varphi \,
\sigma^\mu_j / 2 \right)$ where $\mu \in \{ x, y, z \}$. The
independent coupling of each atom to the electromagnetic vacuum
leads to a decay rate $\Gamma$; the cooperative nature of decay for
atom pairs $i,j$ is reflected by mutual decay rates $\Gamma _{ij}$
(notice that in the following we will use the convention $\Gamma
_{ii} =\Gamma$). A second effect of the collective coupling of the
atoms to the vacuum is the coherent dipole-dipole interaction
characterized by the frequency shifts $\Omega _{ij}$. Both functions
depend on $r_{ij}$ and $\theta$ (as detailed for example in
~\cite{ostermann2012cascaded}). The dynamics of the system can be
described by solving a master equation for $\rho$ (the density
matrix of the whole system of $N$ emitters),
\begin{equation} \label{master}
\frac{\partial \rho} {\partial t}=i[\rho ,H]+\mathcal{L}[\rho ],
\end{equation}
where the Hamiltonian is given by
\begin{equation}
H = \frac{\omega}{2} \sum_{i} \sigma _i^z + \sum_{i \neq j}\Omega _{ij} \, \sigma _i^+ \sigma_j^-
\end{equation}
with $\omega=\omega_0-\omega_l$ ($\omega_l$ is the reference
frequency) while the effect of dissipation is quantified by the
Liouvillian
\begin{equation}
\mathcal{L}[\rho ]= \frac{1}{2} \sum_{i,j} \Gamma _{ij} \left[ 2 \sigma_i^- \rho \, \sigma_j^+ -\sigma_i^+ \sigma_j^- \rho -\rho \, \sigma_i^+ \sigma_j^- \right].
\end{equation}
A typical procedure in spectroscopic experiments with two-level
systems is the Ramsey method of separated oscillatory fields
\cite{Ramsey1990molecula}. The sequence assumes the ensemble of
spins initiated in the ground state at time $t_i$ such that $\langle
S^z\rangle (t=0)= - N/2$ where $S^z= \sum_i \sigma_i^z /2$. Three
stages follow: (i) a first quick pulse between $t_i$ and $t=0$
rotates the atoms into a collective state in the $xy$-plane that
exhibits maximal dipole, (ii) the system evolves freely for the time
$\tau$ and (iii) a second quick pulse flips the spins up. The
detected signal is then a measure of population inversion and
therefore proportional to $\langle S^z \rangle (t_f)$. Analysis of
this signal gives the sensitivity as a figure of merit in metrology
as
\begin{equation}\label{sensdef}
\delta \omega =\min \left[\frac{\Delta S^z (\omega,\tau)}{\left|
\partial_{\omega}\langle S^z\rangle (\omega,\tau) \right| }\right],
\end{equation}
where the minimization is performed with respect to $\omega$.

We follow the dynamics as described above in a density matrix
formalism. We start with an initial density matrix $\rho_i=\left
\vert G \right \rangle \left \langle G \right \vert$, transform it
into $\rho_0 = \mathcal{R}_1 \, \rho_i \, \mathcal{R}^\dagger_1$,
evolve it into $\rho_\tau$ by solving Eq.~(\ref{master}) and finally
transform it into $\rho_f = \mathcal{R}_2 \, \rho_\tau \,
\mathcal{R}^\dagger_2$. The detected signal and its variance are
computed as $\langle S^z \rangle$ and $\Delta S^{z}$ from $\rho_f$.

As a basis of comparison, let us consider the situation of
independent systems ($\Gamma _{ij} = 0$ and $\Omega _{ij} = 0$ for
$i\neq j$). The rotation pulses are $\mathcal{R}_1 = \mathcal{R}_2 =
\bigotimes_j \mathcal{R}^{(j)}_y [\pi/2]$ and the resulting
sensitivity is
\begin{equation} \label{sens-indep}
\left[ \delta\omega \right]_{\text{indep}} = \min
\left[{\frac{\sqrt{e^{\Gamma \tau} - \cos^2(\omega \tau)}}{\sqrt{N}
\left| \tau \cdot \sin(\omega \tau) \right| }}\right]= \frac{\exp
(\Gamma \tau/2)}{\tau \sqrt{N}},
\end{equation}
Further optimization with respect to the interrogation time gives an
optimal $\tau_{opt}=2/\Gamma$ and optimal sensitivity $\Gamma
e/2\sqrt{N}$, which shows that the main impediment of Ramsey
interferometry is the limitation in the interrogation times owing to
the decay of the transition dipoles.

As a principal advance of this paper, we propose a generalized
Ramsey sequence (as illustrated in Fig.~\ref{fig1}) that deviates
from the typical one by a redesign of the two pulses at times $t =
0$ and $t= \tau$, intended to drive the spin system into states that
are protected from the environmental decoherence. To accomplish
this, one complements the normal $\pi/2$ pulse with a phase
distribution pulse, which for a particular atom $j$ is represented
by a rotation around the $z$-direction with the angle
$\varphi_{j}^{(m)}=2\pi m /N (j-1)$, where $m=1,...[N/2]$ and
$[N/2]$ is the smallest integer before $N/2$. The first Ramsey pulse
operator is than
\begin{equation*}
  \mathcal{R}_1 = \, \bigotimes_j \mathcal{R}^{(j)}_z \left[ \varphi_{j}^{(m)} \right] \cdot \mathcal{R}^{(j)}_y \left[ \frac{\pi}{2} \right].
\end{equation*}
The choice of the rotation angles is straightforward to motivate: at
time $t=0$, for any of the angle distributions defined above by the
set of $\varphi_{j}^{(m)}$, the system is in a state of zero average
collective spin. At an intuitive level this means that the system
folds from a state of maximal classical dipole moment to a
non-radiative dipole of zero average. Notice for example that for
small atom-atom separations, collective states of higher symmetry
are shorter lived than the rest of the states; in the Dicke picture
of atoms identically coupled to the vacuum, this culminates in the
maximally superradiant state that exhibits a decay rate $N \Gamma$.
One can eventually deduce the proper rotations that ensure the
asymmetry of the chosen states. This can be derived from the
orthogonality of the initial state $\left| {\psi_\varphi} \right
\rangle = \bigotimes_{j = 1}^{N} \left[ \ket{g} + \left( e^{i
\varphi} \right)^j \ket{e} \right]/\sqrt{2}$ to the multitude of
symmetric states of the system. While generally this is an
unsolvable problem, we can get some insight from the orthogonality
to the symmetric state in the single excitation subspace, the
so-called W-state $\left| {W} \right \rangle$. The imposed
orthogonality $\braket{W | \psi_\varphi} = \sum_{j = 1}^{N} \left(
e^{i \varphi} \right)^j = 0$ leads to the solutions $\varphi=2\pi m
/N$ which we use to build the $\varphi_{j}^{(m)}$.

To prepare the system for population difference detection, at time
$\tau$ the phase spread is reversed and a $\pi/2$ pulse follows
\begin{equation*}
   \mathcal{R}_2 = \bigotimes_j \, \mathcal{R}^{(j)}_y \left[ \frac{\pi}{2} \right] \cdot \mathcal{R}^{(j)}_z \left[- \varphi_{j}^{(m)} \right].
\end{equation*}

\begin{figure}[t]
 \includegraphics[width=0.98\columnwidth]{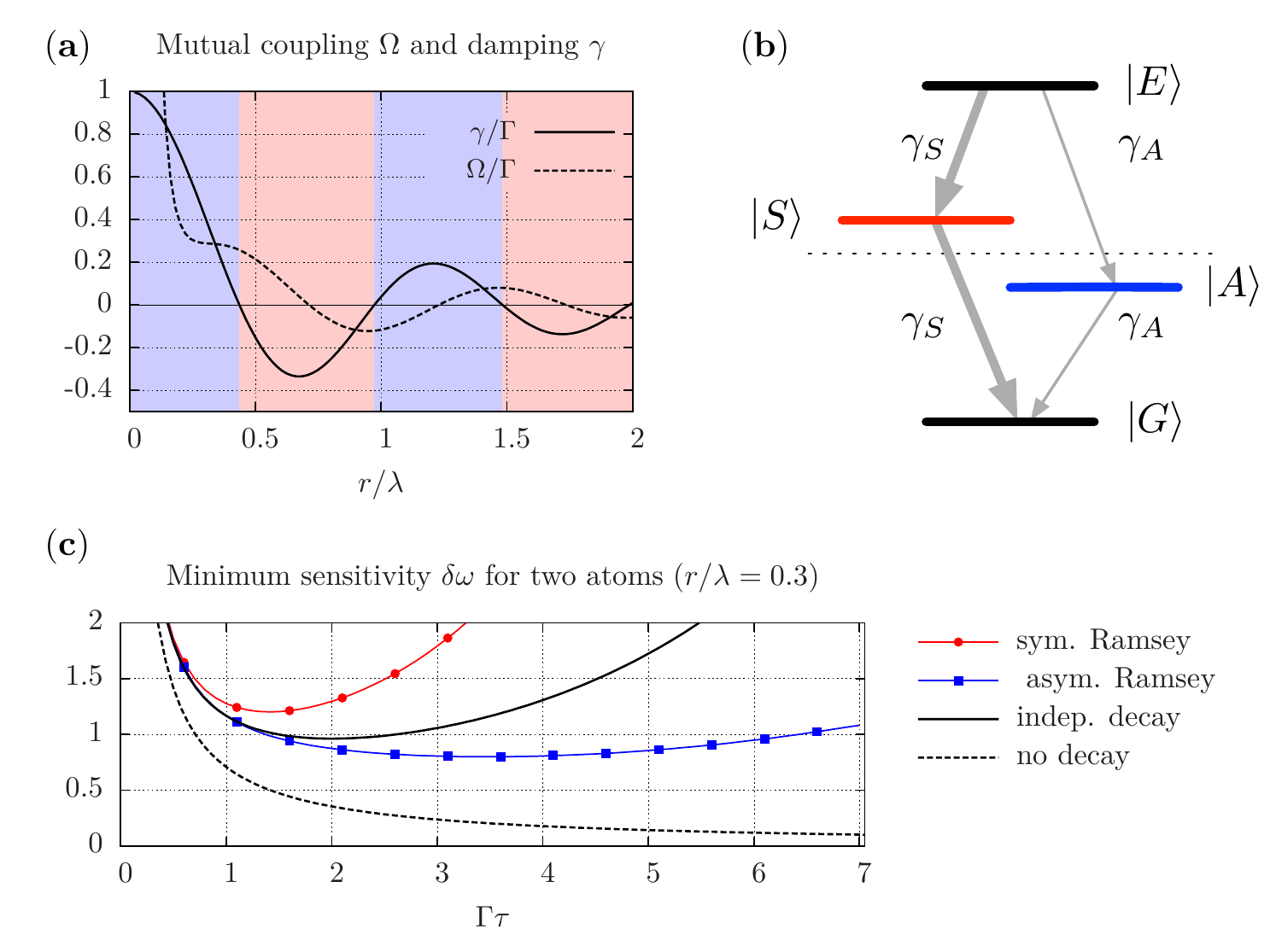}
 \caption{\emph{Two atoms metrology}. (a) Normalized mutual
decay rate and dipole-dipole frequency shift for a pair of atoms as
a function of $r/\lambda$. For positive/negative $\gamma$, the
asymmetric state is subradiant/superradiant (indicated by the
color-coded regions). (b) Level scheme for two interacting atoms in
the collective basis showing the two independent decay channels with
modified rates $\gamma_S$ and $\gamma_A$. (c) The optimal
sensitivity as a function of normalized interrogation time. The atom
separation is $r/\lambda = 0.3$ corresponding to $\gamma \approx
0.41~\Gamma$ and $\Omega \approx 0.29~\Gamma$. The minimum for the
asymmetric addressing is reached around $\tau \simeq 2/\gamma_A$.}
 \label{fig2}
\end{figure}

\textbf{\emph {Analytical results for two atoms}} Analytical results
are easily derived for the case of two atoms and allow us to already
elucidate the differences between typical Ramsey detection and the
asymmetric Ramsey procedure. Let us consider atoms $1$ and $2$
separated by a distance $r$ with a mutual decay rate $\gamma =
\Gamma_{12}(r)$ and dipole-dipole interaction quantified by $\Omega
= \Omega_{12}(r)$ (their dependence on $r$ is shown in
Fig.~\ref{fig2}a). The diagonalization of the Hamiltonian is
performed by a transformation from the bare basis $ \left \lbrace
\left\vert gg \right\rangle, \, \left\vert ge \right\rangle, \,
\left\vert eg \right\rangle, \, \left\vert ee \right\rangle \right
\rbrace$ to the collective basis $ \left \lbrace \left\vert G
\right\rangle, \, \left\vert S \right\rangle, \,\left\vert A
\right\rangle, \, \left\vert E \right\rangle \right \rbrace$ with
$\left\vert G \right\rangle = \left\vert gg \right\rangle$,
$\left\vert S \right\rangle = \left( \left\vert eg
\right\rangle+\left\vert ge \right\rangle \right)/\sqrt{2}$,
$\left\vert A \right\rangle = \left( \left\vert eg
\right\rangle-\left\vert ge \right\rangle \right)/\sqrt{2}$ and
$\left\vert E \right\rangle = \left\vert ee \right\rangle$. This
change of basis diagonalizes the dissipative dynamics as well, and
leads to two independent decay channels with damping rates
$\gamma_{S}=\Gamma+\gamma$ and $\gamma_{A}=\Gamma-\gamma$ as
illustrated in Fig.~\ref{fig2}b.

We follow the evolution of the initially prepared $\rho_i=\left
\vert G \right \rangle \left \langle G \right \vert$ in the
collective basis and compute the detected signal and its variance
from the density matrix at time $\tau$. For the symmetric Ramsey
sequence one obtains $\langle S^z \rangle_{\text{S}} = 2\sqrt{2} \Re
\left( {\rho^{ES}_{\tau}+\rho^{SG}_{\tau}}\right)$ which can be
calculated by solving the evolution between $0$ and $\tau$ from the
following set of coupled equations
\begin{subequations}
\begin{align}
\dot \rho^{ES} =& \left[ -\frac{2\Gamma+\gamma_S}{2} - i (\omega-\Omega) \right] \rho^{ES},\\
\dot \rho^{SG} =& \left[ -\frac{\gamma_S}{2} - i(\omega+\Omega) \right] \rho^{SG} + \gamma_S \, \rho^{ES}.
\end{align}
\end{subequations}

\begin{figure*}[t]
 \includegraphics[width=1.99\columnwidth]{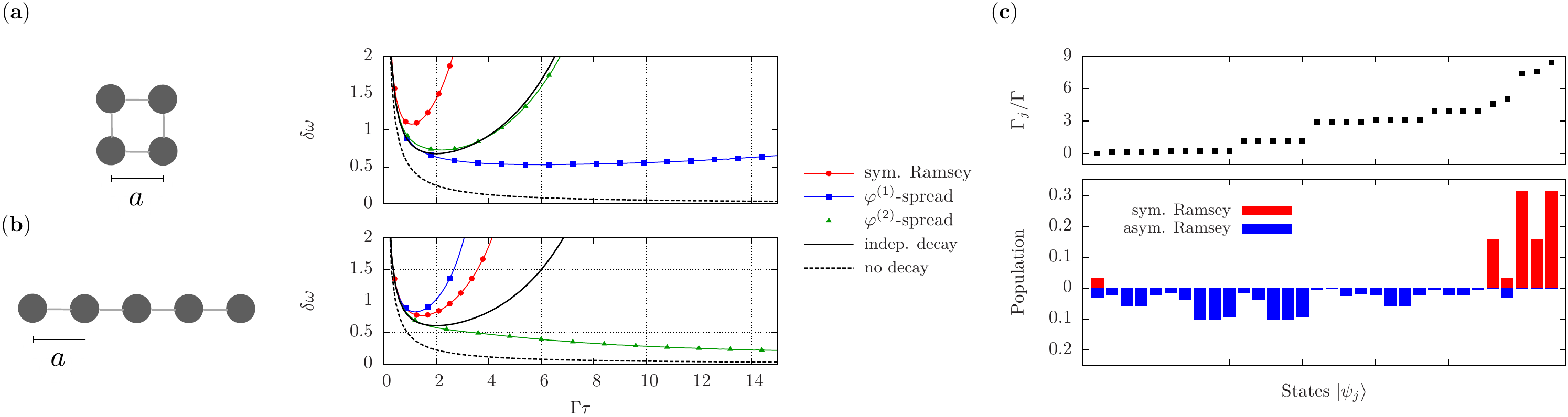}
 \caption{\emph{Numerical investigations}. a) and b) Numerical results for the square and
a 5 atom-chain. c) Results of diagonalization for an ideal system of
5 equally mutually coupled emitters. The states are ordered with
increasing effective decay rate in the top third of the panel. The
occupancy of the $32$ states is shown in red(middle part)/blue(lower
part) for symmetric vs. asymmetric Ramsey sequences.  The two
sequences excite states of very different behavior: superradiant vs.
subradiant.}
 \label{fig3}
 \end{figure*}

The computation of the signal variance requires the derivation of
$\left \langle \left( S^z \right)^2 \right \rangle_{\text{S}} = 2
\left[1+\rho^{SS}_{\tau}-\rho^{AA}_{\tau}+2 \Re \left(
\rho^{EG}_{\tau} \right) \right]$ thus solving
\begin{subequations}
\begin{align}
\dot \rho^{EE} =& -2 \Gamma \rho^{EE},\\
\dot \rho^{SS}=& -\gamma_S \left( \rho^{SS}-\rho^{EE} \right),\\
\dot \rho^{AA} =& -\gamma_A \left( \rho^{AA}-\rho^{EE} \right),\\
\dot{\rho}^{EG} =&-(\Gamma+2i \omega) \rho^{EG}.
\end{align}
\end{subequations}
In contrast, for the asymmetric Ramsey sequence we get $\langle S^z
\rangle_{\text{A}} = 2\sqrt{2} \Re \left( {\rho^{EA}_{\tau} -
\rho^{AG}_{\tau}}\right)$ and $\left \langle \left( S^z \right)^2
\right \rangle_{\text{A}} = 2
\left[1+\rho^{AA}_{\tau}-\rho^{SS}_{\tau} - 2 \Re \left(
\rho^{EG}_{\tau} \right) \right]$, where the extra coherences can be
derived from the solutions of
\begin{subequations}
\begin{align}
\dot \rho^{EA} =& \left[ -\frac{2\Gamma+\gamma_A}{2} - i (\omega + \Omega) \right] \rho^{EA},\\
\dot \rho^{AG} =& \left[ -\frac{\gamma_A}{2} - i(\omega - \Omega) \right] \rho^{AG} + \gamma_A \, \rho^{EA}.
\end{align}
\end{subequations}
The minimum sensitivities depending on $\tau$ after optimization
with respect to $\omega$ can be expressed as
\begin{subequations}
\begin{align}
\left[ \delta \omega \right]_{\text{S}} =& \frac{\sqrt{ 2 \left( 1 + a_S e^{-2 \Gamma \tau}+b_S e^{-\gamma_S \tau} - c_S e^{-\gamma_A \tau} \right)}}{\tau \cdot e^{-\gamma_S \tau/2} \left( e^{-\Gamma \tau} \mathcal{A}^-_S + \mathcal{A}_S^+ \right)} \\
\left[ \delta \omega \right]_{\text{A}} =& \frac{\sqrt{ 2 \left( 1 +
a_A e^{-2 \Gamma \tau}+b_A e^{-\gamma_A \tau} - c_A e^{-\gamma_S
\tau} \right)}}{\tau \cdot e^{-\gamma_A \tau/2} \left( e^{-\Gamma
\tau} \mathcal{A}_A^+ + \mathcal{A}^-_A \right)},
\end{align}
\end{subequations}
where $a$, $b$, $c$ and $\mathcal{A}^\pm$ are given by the system's
geometry and are listed in the Appendix. While the above expressions
are tedious, simplifications are possible in the limit of large
$\tau$. Assuming a separation of timescales for example when
$\gamma_A\ll\Gamma,\gamma_S$, the sensitivity $\left[ \delta \omega
\right]_{\text{A}}$ scales similarly to the independent sensitivity
of Eq.~(\ref{sens-indep}) with $\Gamma$ replaced by $\gamma_A$. This
actually holds approximately even in the intermediate regime shown
Fig.~\ref{fig2}c where $\gamma_A\simeq 0.59~\Gamma$, as transpiring
from the scaling of the blue (squares) line. For closely spaced
atoms, the result is easy to interpret and extremely encouraging
since it allows for large interrogation times and direct improvement
of the minimum sensitivity. In the general case, of varying the
distance between atoms for example to the second region of
Fig.~\ref{fig2}a, the symmetric state becomes subradiant instead and
the symmetric procedure is the optimal one, however providing only a
minimal gain over the independent atom case. This is relevant for
the case of linear atom chains separated by a magic wavelength
~\cite{takamoto2003spectroscopy}, where the conventional Ramsey
technique is optimal.

\textbf{\emph {Numerical results for several atoms}} Let us now
extend our model to more general configurations of a few two-level
systems in various geometries. In principle, the configuration can
be generalized to a 2D or 3D lattice but one ends up with large
Hilbert spaces rather quickly that render simple numerical methods
unfeasible. To illustrate the effectiveness of the asymmetric Ramsey
method we particularize to the two situations depicted in Fig.~
\ref{fig3}, i.e., square and linear geometries. The results are
presented in Fig.~\ref{fig3}a,b for all possible phase-spread angle
sets, i.e., varying the index $m$ of $\varphi_{j}^{(m)}$ from 1 to
$[N/2]$ ($N=4$ for square and 5 for the chain) and for a lattice
constant $a/\lambda=0.30$.

To provide a simplified general understanding of the results shown
in Fig.~\ref{fig3} let us first present a numerical analysis of a
simplified case of 5 atoms equally coupled to each other. To this
end we shall point out that the method we present should be quite
general and applicable to similar systems where the naturally
occurring electromagnetic bath that provides mutual decay channels
as well as dipole-dipole interactions for dense ensembles of quantum
emitters is replaced, for example, by the common interaction of
atoms with a decaying optical cavity field \cite{holland2010}. In
such a case, by tailoring the atom-field interaction, one can
simulate a reservoir that leads to equal mutual coupling between any
pair of atoms and equal dipole-dipole couplings. Simultaneous
diagonalization of the Hamiltonian and Liouvillian is then possible
that leads to $2^N$ states $|\phi_j\rangle$ each with an associated
decay channel $\Gamma_j$. In Fig.~\ref{fig3}c, we show this exact
diagonalization for $N=5$ and associated $\Gamma_j/\Gamma$ arranged
in increasing magnitude from left to right. The upper (red)
histogram shows the population distribution for a Ramsey operation
while the blue (lower) histogram provides the comparison with the
asymmetric Ramsey excitation scheme. The conclusion is
straightforward in that it shows that the conventional Ramsey
technique excites on average states decaying faster than $\Gamma$
and the asymmetric scheme populates subradiant states.

While the examples studied in Fig.~\ref{fig3} are a
proof-of-principle for the phase-spread mechanism we propose, a
general optimization for arbitrary distances and geometries is not
straightforward and needs to be accompanied by more sophisticated
numerical simulations. For example, in the linear chain case for a
ratio $a/\lambda=0.15$, the two nearest neighbors contribute
positively while the outer ones feature a negative coupling (see
Fig.~\ref{fig2}a). The strategy to be employed is therefore not
clear since the various phase shifts are distributed differently
along the chain.  For example, as seen in Fig.~\ref{fig3}b, a simple
$\varphi_{j}^{(1)}$ phase distribution performs worse than the
symmetric Ramsey sequence while great improvement is introduced by
applying $\varphi_{j}^{(2)}$ shifts.

Experimental investigations of the mechanism described above must
mainly address the question of individual phase writing on
distinguishable emitters. As one particular realization, a chain of
atoms excited by a laser tilted by some angle $\alpha$ opens up the
possibility of imprinting a varying phase $\varphi _j = k_0 (j-1) a
/ \cos (\alpha)$ for ´the $j^{th}$ atom. To realize an optimal
phase-spread by angles $\varphi_j^{(1)}$ for example, one has to
fulfill $\alpha = \arccos \left(Na/\lambda_0 \right)$. Note, that
interestingly for a strontium magic wavelength lattice, excitation
at about $ 90^{\circ}$ automatically excites long lived exciton
states close to the optimum. In a 2D lattice this still is fulfilled
quite well by excitation from the third direction perpendicular to
the plane. For a cube the situation is more tricky and requires
careful angle optimization for which preliminary calculations are
promising and will be fully investigated in a future publication.

Dipole-dipole interactions and collective decay also play a major
role in recent experiments of several superconducting q-bits coupled
to CPW transmission lines and
resonators~\cite{wallraff2013private,lalumiere2012cooperative,lalumiere2013input}.
Here, on the one hand the distance of the particles is much smaller
than a wavelength so the effects are very large, but on the other
hand the individual transition frequencies, Rabi amplitudes and
phases can be controlled very well. The situation is similar to the
above described engineered bath for atoms in an optical cavity where
the cavity field dissipation induces non-local collective decay of
atomic states. In both cases, the individual atoms cane be
separately addressed: for example a tunable magnetic field gradient
can provide the necessary phase gradient across the ensemble
allowing for the asymmetric Ramsey procedure to be tested.

Let us finally remark on the connection of our scheme to
multipartite entanglement. The folding of the collective atomic
state to a partially protected subspace that suffers less from the
effects of decoherence hints towards the possibility of preparing
entangled atomic states via dissipative techniques. More concretely,
assuming the ideal case discussed above where the Liouvillian can be
diagonalized and the dissipative evolution consists of independent
channels of decay, and assuming the system in a state of minimal
dissipation, the final state of the system after considerable
evolution time $\tau$ will be an eigenstate of the Hamiltonian with
probable quantum correlations (as a basis of comparison consider the
state $|A\rangle$ and its entanglement for the 2 atom case). An
optimization of our scheme by introducing a proper rotation of this
final state before detection seems therefore feasible. Moreover,
protection of collective states can as well be relevant to schemes
where generated entanglement (such as spin squeezed state generation
via one axis or two-axis twisting) is exploited and where
decoherence has a extremely fast destructive effect. Application of
a general principle that would allow multiparticle entanglement to
be mapped from fast decaying state to subradiant states could be of
great interest.

\textbf{\emph {Concluding remarks.}} We have shown that quantum
metrology applications such as frequency measurements via the Ramsey
method can benefit from a state protective mechanism that can be
directly connected to a transformation that folds initial collective
states from the surface of the Bloch sphere to its center. While we
have mainly analyzed the simplest collective bath where the vacuum
mediates interactions among closely spaced quantum emitters, the
procedure should be quite general as for example in the case of
engineered baths (atoms in mode-structuring cavities,
superconducting q-bits coupled to CPW transmission lines). The
connection and application of the mechanism to multiparticle
entangled systems indicates possible future directions as it hints
towards investigation in i) dissipation-induced entanglement and ii)
entanglement (spin squeezed states) protection mechanism.

\textbf{\emph {Acknowledgements}} We are grateful to H. Zoubi, M.
Skotiniotis, W. Niedenzu and M. Holland for useful comments on the
manuscripta and to S. Kr\"{a}mer for assistance with numerics. We
acknowledge the use of the QuTiP open-source
software~\cite{QuTip2013} to generate Fig.~\ref{fig1}. Support has
been received from DARPA through the QUASAR project (L.~O. and
H.~R.) and from the Austrian Science Fund (FWF) via project
P24968-N27 (C.~G.).

\newpage

\textbf{\emph {Appendix}} The $a, b, c, \mathcal{A}^\pm$ depend on
the relative distance between atoms $r$ via the $\gamma_{S,A}$ decay
rates
\[ \begin{aligned}
a_S= \frac{1}{4} \left( \frac{\gamma_A}{\gamma_S} - \frac{\gamma_S}{\gamma_A} \right) \qquad& a_A = \frac{1}{4} \left( \frac{\gamma_S}{\gamma_A} - \frac{\gamma_A}{\gamma_S} \right)\\
b _S= \frac{4 \Gamma - \gamma_S}{4 \gamma_A} \qquad&  b_A = \frac{4 \Gamma - \gamma_A}{4 \gamma_S} \\
c _S= \frac{\gamma_A}{4 \gamma_S} \qquad& c _A= \frac{\gamma_S}{4 \gamma_A} \\
\alpha^\pm _S= 1 \pm \frac{\Gamma \gamma_S}{\Gamma^2+4 \Omega^2} \qquad& \alpha^\pm _A= 1 \pm \frac{\Gamma \gamma_A}{\Gamma^2+4 \Omega^2} \\
B_S = \frac{2 \Omega \gamma_S}{\Gamma^2 + 4 \Omega^2} \qquad& B_A = \frac{2 \Omega \gamma_A}{\Gamma^2 + 4 \Omega^2} \\
A_S^\pm = \sqrt{\left( \alpha^\pm_S \right)^2 + B_S^2} \qquad&
A_A^\pm = \sqrt{\left( \alpha_A^\pm \right)^2 + B_A^2}
\end{aligned} \]

\end{document}